\documentclass[]{interact}

\usepackage[T1]{fontenc}
\usepackage[utf8]{inputenc}
\usepackage{epstopdf}
\usepackage[numbers,sort&compress]{natbib}
\bibpunct[, ]{[}{]}{,}{n}{,}{,}

\usepackage{tabularx}
\usepackage{array}
\usepackage{xurl}
\usepackage[hidelinks]{hyperref}
\usepackage{textcomp}
\usepackage{xcolor}
\begin{document}
\jname{High Pressure Research}
\articletype{Research Article}
\title{SeriesXRD: automated reflection tracking and phase identification in high-pressure X-ray diffraction series}
\author{%
\name{Rush Maples\textsuperscript{a,b}, Zackary Rehfuss\textsuperscript{b}, Hongyi Liu\textsuperscript{b} and Sheng Ran\textsuperscript{b}}
\affil{\textsuperscript{a}School of Physics, Georgia Institute of Technology, Atlanta, Georgia, USA; \\
\textsuperscript{b}Department of Physics, Washington University in St. Louis, St. Louis, Missouri, USA}%
}
% TODO before submission: add \thanks{CONTACT... Email:...} after the corresponding author's name.
\maketitle

 \begin{abstract}
 High-pressure synchrotron X-ray diffraction increasingly produces hundreds to thousands of detector frames in a single experiment, while separating, assigning and tracking weak reflections across an entire series often remain manual. This bottleneck is particularly severe for micron- and nanoscale samples in diamond-anvil cells, where scattering from the anvils, gasket, pressure medium and pressure markers can dominate the sample signal. We present SeriesXRD, an open-source workflow that analyzes diffraction as a series rather than as independent patterns. The program suppresses azimuthally localized diamond scattering, subtracts background, fits and tracks reflections between frames, matches known phases using pressure-dependent lattice parameters, subtracts confidently assigned contributions, and groups persistent unexplained reflections for follow-up as candidate sample-associated or unknown-phase features. We demonstrate the method on 1288 frames collected from a \(\sim30\)-nm-thick layered UOTe flake in a diamond-anvil cell over 3.5--50.7~GPa. The dataset is strongly background dominated, with detectable UOTe diffraction in only approximately 4\% of the frames. SeriesXRD recovered the sample-associated reflections identified in the original manual analysis while reducing the corresponding phase-assignment and residual-analysis workflow from more than two months of researcher time to an overnight automated run. Across the full series, 17\,143 peaks were fitted, 14\,395 passed quality criteria, and 90\% of accepted peaks were assigned to known background or reference phases. SeriesXRD therefore converts large, background-dominated diffraction series into a tractable set of phase assignments and coherent unexplained reflections while retaining a traceable record of each decision.
\end{abstract}

\begin{keywords}
high-pressure diffraction; diamond anvil cell; automated phase identification; diffraction series; background subtraction; open-source software
\end{keywords}

\section{Introduction}
Third- and fourth-generation synchrotron sources make it straightforward to collect diffraction series much faster than they can be analyzed frame by frame. High-pressure measurements are a particularly demanding example. A diamond-anvil-cell experiment may acquire diffraction patterns at many pressures and at many spatial positions within the sample chamber, with weak sample reflections superimposed on much stronger scattering from the diamond anvils, gasket, pressure medium and pressure markers. The problem becomes more severe as sample dimensions shrink: the beam can traverse millimetres of diamond and micrometres of gasket or pressure medium while interacting with only tens of nanometres of the material of interest. In such measurements, separating sample-associated reflections from nonsample contributions and following their evolution across the series can require as much effort as the subsequent structural analysis.

 Existing open-source tools solve important parts of the diffraction workflow. pyFAI performs detector calibration and azimuthal integration \cite{ashiotis2015}; DIOPTAS provides interactive calibration, masking, integration and comparison with candidate phases \cite{prescher2015}; and GSAS-II supports structural and sequential Rietveld refinement once the relevant phases are established \cite{toby2013}. These tools are effective for individual patterns or for a series in which the user already knows which phases contribute to each frame. The remaining bottleneck is the series-level analysis: separating cell and marker scattering from the sample, fitting and tracking reflections as pressure changes, assigning known phases within physically plausible pressure ranges, and recognizing weak or unexpected peaks that remain after known contributions are removed.

 This decision process is still commonly performed by visual inspection. For a small number of patterns that approach is practical. For spatial maps or pressure series containing hundreds to thousands of frames, however, repeated manual masking and phase comparison can require weeks to months and is difficult to reproduce exactly. It also creates a risk that weak sample peaks are overlooked when they occur near intense diamond, gasket or pressure-marker reflections.

 SeriesXRD was developed to automate this intermediate stage. Rather than treating each pattern independently, it uses the continuity of a diffraction series: peaks are fitted and tracked from frame to frame, known phases are compared at physically allowed pressures, confidently assigned contributions are removed, and the residual is searched again for weaker features. Unexplained reflections that recur and move together through the series are grouped as candidate sample or unknown-phase features. The goal is not to replace crystallographic refinement or to claim a structure from unassigned peaks. It is to convert a large experimental series into a traceable set of phase assignments and coherent residual reflections that require scientific interpretation, while preserving the evidence behind each decision.

 We validate this approach on an intentionally difficult high-pressure dataset from an approximately 30~nm thick flake of the layered material UOTe. The X-ray beam was incident on the \(ab\) plane of the flake, and the weak sample reflections were measured in the presence of scattering from approximately 2~mm of diamond on each side of the cell and approximately 10~\(\mu\)m of pressure medium. The experiment therefore provides a stringent test of the intended use case: separating known pressure-cell and marker contributions, tracking sample-associated reflections across pressure, and recovering coherent residual features for further interpretation. The automated result is compared directly with the prior manual analysis of the same experiment.

 \begin{figure}[t]
\centering
\includegraphics[width=.8\textwidth]{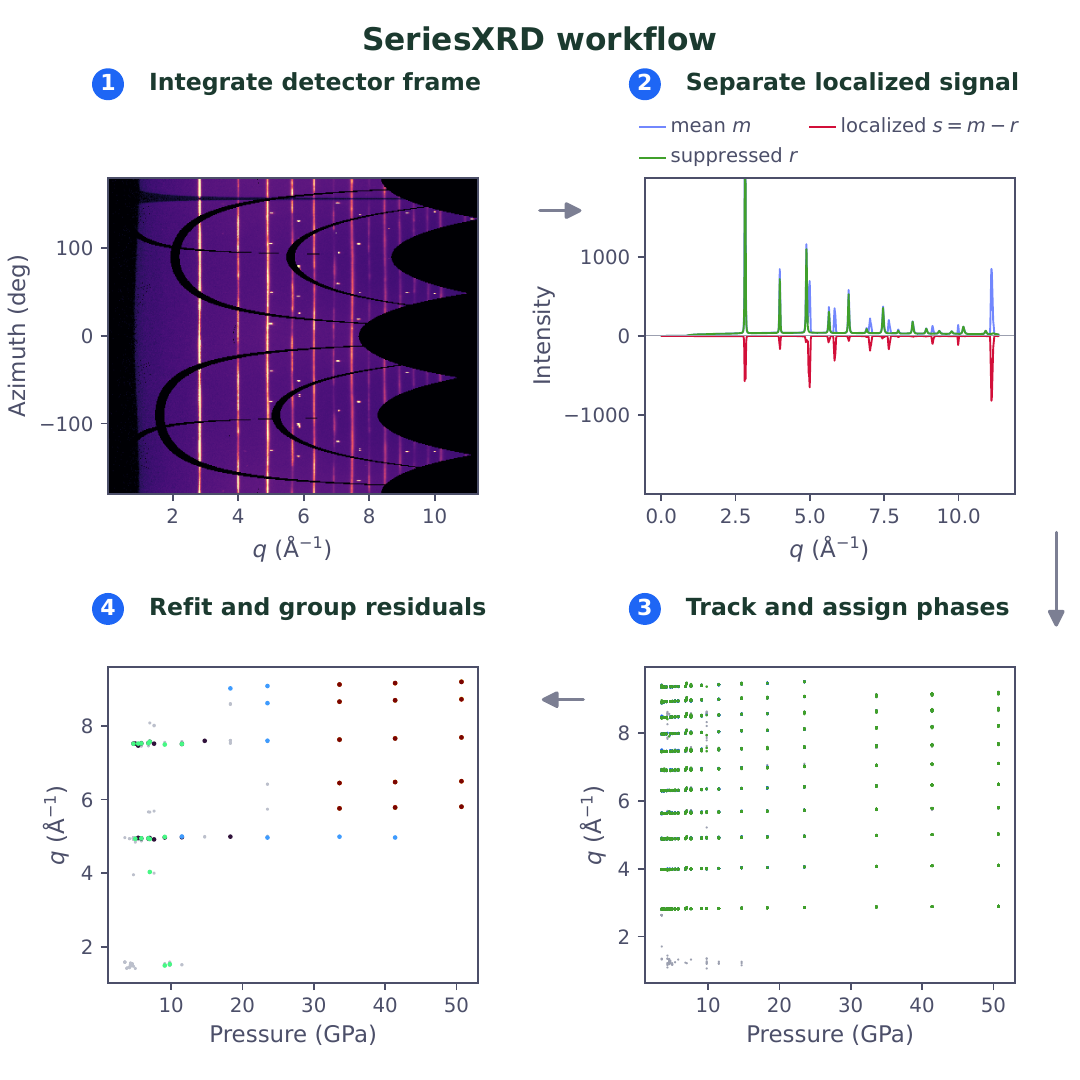}
\caption{SeriesXRD workflow for a high-pressure diffraction series. A detector frame is azimuthally integrated into complementary channels that preserve powder-like scattering while isolating azimuthally localized intensity. After baseline subtraction, peaks are fitted and tracked between frames. Known phases are tested at pressures consistent with the experimental metadata and are subtracted only after passing assignment criteria. The residual is then re-fitted so that weak sample-associated or otherwise unexplained reflections can be tracked across the series.}
\label{fig:workflow}
\end{figure}

 \section{SeriesXRD workflow}
\subsection{Reduction and separation of localized scattering}
SeriesXRD accepts integrated patterns or performs azimuthal integration through pyFAI \cite{ashiotis2015}. For diamond-anvil-cell data, the reduction stage retains both the azimuthal mean and a spot-suppressed channel. This distinction is useful because powder diffraction is distributed around a Debye--Scherrer ring, whereas single-crystal reflections from the diamond anvils are localized over a narrow azimuthal range and can dominate a conventional azimuthal mean.

Let \(m\) denote the azimuthal mean and \(r\) the spot-suppressed channel. Their difference,
\begin{equation}
 s=m-r,
\end{equation}
 provides a measure of azimuthally localized scattering. A slowly varying baseline \(b\) is estimated from the spot-suppressed pattern using the SNIP algorithm \cite{ryan1988,morhac1997}, yielding the background-corrected signal
\begin{equation}
 c=r-b.
\end{equation}
The localized residual \(s\) is used to flag frames dominated by intense anvil reflections, while \(c\) is used for peak finding when the sample diffraction survives spot suppression. Because strongly textured or coarse-grained samples can themselves produce localized diffraction, the program also compares the mean and spot-suppressed channels and can revert to a less aggressive channel when suppression removes substantial sample intensity. This prevents the diamond-rejection step from being treated as an assumption that all localized intensity is contamination.

 \begin{figure}[t]
\centering
\includegraphics[width=0.62\textwidth]{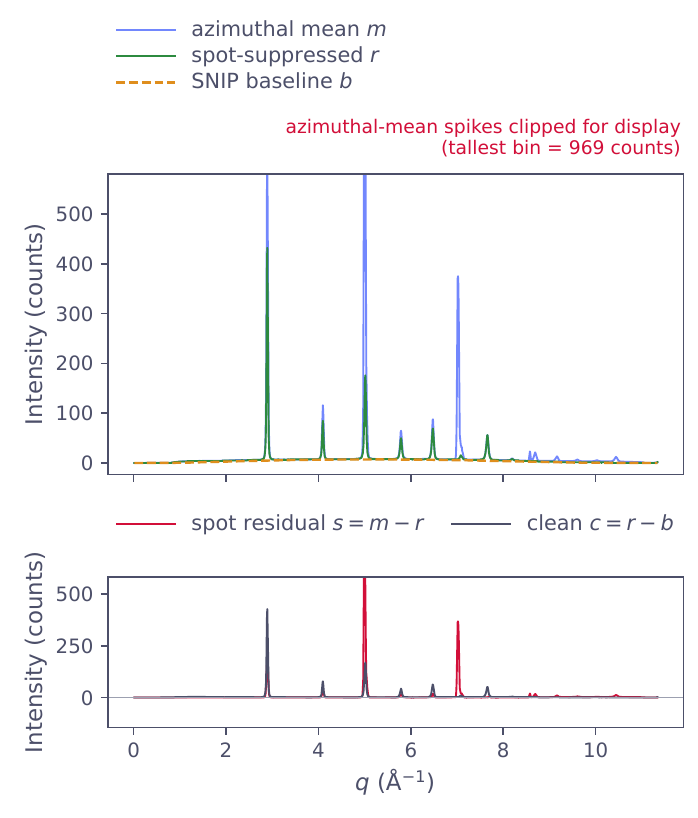}
\caption{Separation of intense localized scattering from the background-corrected diffraction signal in a representative high-pressure frame. The azimuthal mean \(m\) contains strong localized reflections that are suppressed in \(r\); their difference \(s=m-r\) isolates these contributions, while subtraction of the SNIP baseline \(b\) gives the cleaned channel \(c=r-b\) used for peak analysis. The purpose of the separation is not to classify every localized reflection as diamond, but to prevent intense cell scattering from dominating weak sample features.}
\label{fig:separation}
\end{figure}

\subsection{ Peak fitting and tracking across a series }
Peaks are detected relative to a noise estimate based on the median absolute deviation and are fitted with pseudo-Voigt profiles. Overlapping fitting windows are treated jointly. Fits that fail amplitude, convergence, position, width or goodness-of-fit criteria are retained with quality flags rather than silently discarded. The resulting peak positions and their fitted uncertainties are propagated to the phase-matching stage.

 The key distinction from independent frame-by-frame fitting is continuity. A peak identified in one frame seeds the search in the next, so a reflection can retain its track as pressure, position or another experimental variable changes. The matching tolerance scales with the fitted peak width and uncertainty rather than using a single fixed bin width. This series-resolved treatment both reduces missed detections and provides the trajectories needed to distinguish persistent sample reflections from isolated artifacts.

\subsection{ Pressure-aware phase matching and subtraction }
 Known phases are supplied through a phase library containing lattice parameters and, where available, an equation of state. Third-order Birch--Murnaghan \cite{birch1947} and Vinet \cite{vinet1987} forms are supported, together with anisotropic axis-dependent compression. For each frame, SeriesXRD predicts the reflection positions of a candidate phase over an allowed pressure interval and performs one-to-one matching between predicted and observed peaks. One measured peak therefore cannot satisfy several predicted reflections simultaneously. Position uncertainty controls the matching tolerance, while relative intensity can be included only as a weak factor because preferred orientation and limited grain statistics can strongly alter intensities in diamond-anvil-cell measurements.

 Experimental pressure metadata constrains the search. The pressure may be read from filenames, a metadata table or NeXus fields; when direct pressure information is absent, a pressure-marker phase can first be fitted and used to constrain the remaining candidates. This pressure prior is important because diffraction patterns can contain accidental positional coincidences that would otherwise allow a phase to be matched at an unphysical pressure.

A phase is removed from the pattern only after meeting minimum match and confidence criteria. Its fitted contribution is reconstructed and subtracted, after which the residual is passed through the peak-fitting stage again. This second fit is central to the workflow: weak reflections that were unresolved beneath strong gasket, marker or other known peaks can become detectable only after the dominant contribution has been removed.

 Candidate ranking is available to reduce the number of library phases passed to the full matcher. The ranking stage is an acceleration step rather than the scientific basis of the assignment; final calls are made by the pressure-constrained, one-to-one reflection matching described above. Additional synthetic and labelled-pattern benchmarks of the ranker are provided with the software and Supplementary Material.

\subsection{ Unexplained-reflection tracking }
 After known phases have been subtracted, SeriesXRD links residual peaks between neighboring frames and groups tracks that appear, disappear or shift together. The output is a set of recurring \(d\)-spacing fingerprints and the frames in which those fingerprints are present. These groups are intentionally reported as \emph{candidate} sample or unknown-phase features, not as crystallographic identifications. Their purpose is to direct attention to coherent unexplained scattering that would otherwise have to be found by inspecting every frame manually.

 This distinction is particularly important for sparse measurements. If most positions in a spatial scan contain only diamond, gasket, pressure medium and marker scattering, removal of those known contributions leaves a compact criterion for identifying the few positions that contain reproducible additional reflections. The same logic can be applied to pressure or time series, where the appearance or disappearance of a group of residual tracks identifies the portion of the dataset that merits detailed structural analysis.

\subsection{ Relationship to existing diffraction tools}
SeriesXRD is designed to sit between integration and refinement rather than replace either. It calls pyFAI for integration, complements the interactive inspection available in DIOPTAS, and exports identified patterns and phases for subsequent GSAS-II refinement \cite{ashiotis2015,prescher2015,toby2013}. Quantitative phase fractions and final structural parameters should therefore still be obtained by an appropriate refinement procedure. SeriesXRD addresses a different task: deciding, reproducibly and at series scale, which frames contain useful signal, which reflections can be explained by known components of the pressure cell, and which reflections remain for scientific interpretation.

 \section{Validation on a sparse ultrathin-sample dataset}
 \subsection{ Experimental test}
The end-to-end test comprises 1288 detector frames collected at the Advanced Photon Source with wavelength \(\lambda=0.3066\)~\AA. The series spans 22 pressures from 3.5 to 50.7~GPa and 56 spatial positions in the sample chamber. Pressure was measured independently by ruby fluorescence and carried into the analysis as frame metadata. The specimen was an approximately 30~nm thick UOTe flake measured with the incident beam along the crystallographic \(c\) direction, i.e. incident on the \(ab\) plane. The pressure medium was neon, and tungsten contributions arose from the pressure marker and gasket.

 This dataset was selected because it represents the regime in which manual diffraction screening is most costly. The sample is orders of magnitude thinner than the X-ray path through the diamond anvils and pressure medium, and the spatial scan contains sample diffraction only near the flake. The majority of frames therefore consist predominantly of cell, marker and medium scattering. In the original analysis, the full series was inspected manually to mask or identify tungsten, ruby/cell contributions and the sample-bearing patterns. That process required more than two months of researcher time. SeriesXRD performs the corresponding series-level screening and phase subtraction in an overnight automated run.

 The present count is approximately 48 of 1288 frames, or about 4\%. This sparsity is the central challenge: more than 95\% of the dataset can be uninformative for the sample while still containing strong diffraction from other components of the pressure assembly.

 \subsection{Agreement with manual analysis}
\begin{figure}[t]
\centering
\includegraphics[width=0.47\textwidth]{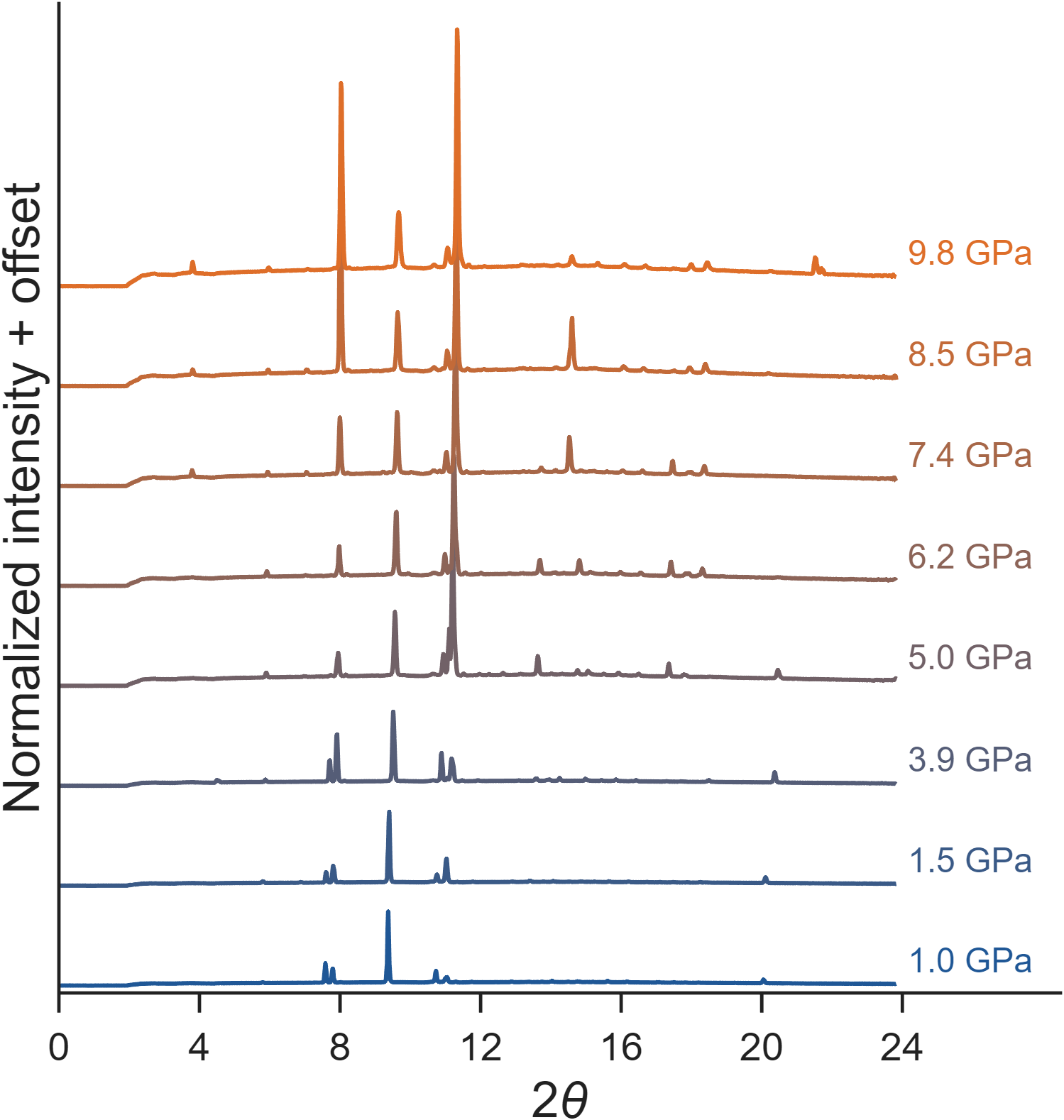}\hfill
\includegraphics[width=0.47\textwidth]{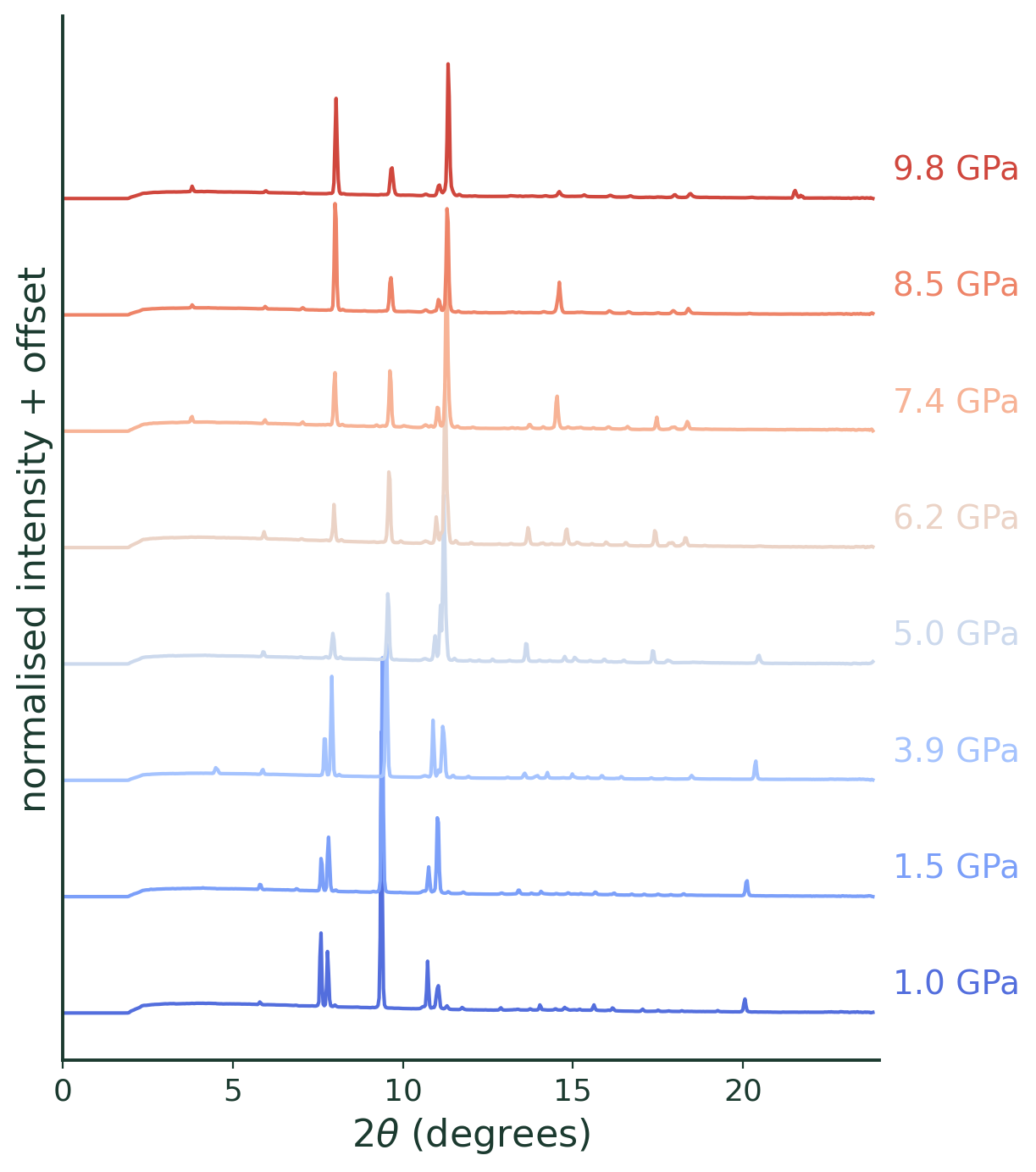}
\caption{Manual and automated analysis of a 28-frame UOTe subset at eight pressures from 1.0 to 9.8~GPa. The manually inspected patterns (left) and SeriesXRD output (right) identify the same sample reflections at the same pressures. At low pressure the diffraction contains reflections associated with a monoclinically distorted, supermodulated structure; below 10~GPa these evolve into a monoclinic structure with restored translational symmetry. Detailed structural interpretation is reserved for the accompanying UOTe study.}
\label{fig:manualauto}
\end{figure}

The automated analysis recovered the same sample-bearing frames as the prior manual inspection. A smaller 28-frame subset at eight pressures from 1.0 to 9.8~GPa provides a direct visual comparison between the manual and automated analyses (Figure~\ref{fig:manualauto}). The same sample reflections are identified at the same pressures in the two treatments, but the SeriesXRD result is generated in a single batch analysis rather than by inspecting and masking the frames individually.

The low-pressure subset also establishes that the residual features are physically meaningful rather than arbitrary unassigned peaks. At low pressure, the UOTe diffraction contains reflections associated with a monoclinically distorted, supermodulated structure. With increasing pressure below 10~GPa, these reflections evolve into those of a monoclinic structure with restored translational symmetry. The full pressure-induced structural evolution is outside the scope of the present methods paper and will be reported separately; here, the comparison is used only to test whether the automated workflow recovers the same experimentally relevant reflections as manual analysis.

 \subsection{Removal of known phases and recovery of weak reflections}
Across the 1288-frame series, SeriesXRD fitted 17\,143 peak profiles, of which 14\,395 (84.0\%) passed all quality checks. Five library entries were tested against each frame. Subtraction of confidently identified phases removed 12\,958 accepted peaks, corresponding to 90.0\% of the accepted first-pass peaks. Of these assignments, 6626 were attributed to the tungsten pressure marker and 6330 to tungsten from the gasket. The remaining signal was then re-fitted rather than treated as a final residual.

Re-fitting after subtraction recovered 1569 additional peaks that had been obscured by the dominant known phases. Combined with the 1437 peaks left unexplained in the first pass, 3006 residual features entered the tracking and clustering stage. These were linked into 187 tracks and grouped into 104 co-occurring clusters, of which 22 contained at least three tracks. The numerical value of a cluster is not that it constitutes a solved phase; it is that a coherent set of reflections can be surfaced automatically from thousands of otherwise background-dominated measurements.

\begin{figure}[t!]
\centering
\includegraphics[trim={0, 1.5cm, 0, 0}, clip, width=\textwidth]{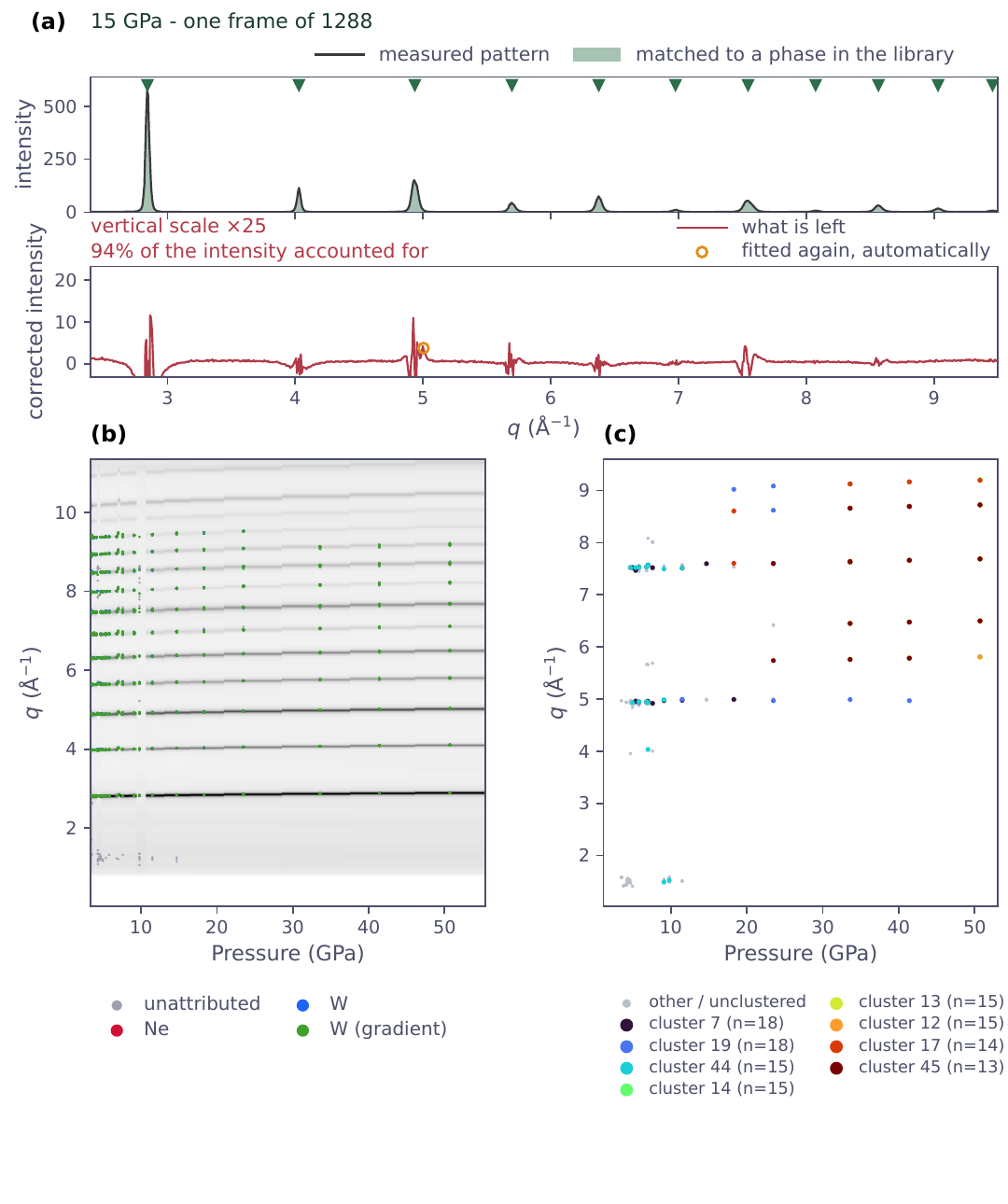}
\caption{Series-level reduction of a background-dominated high-pressure
dataset. (a) Representative pattern before and after subtraction of
confidently assigned known phases; re-fitting the residual recovers weak
reflections that overlap stronger contributions. (b) Full-series map showing
fitted reflections assigned to known phases and the remaining unattributed
features. (c) Residual reflections linked into tracks and grouped by
co-occurrence. The residual groups are candidate sample or unknown-phase
features and are not, by themselves, crystallographic identifications.}
\label{fig:seriesmap}
\end{figure}

 \subsection{Pressure consistency}
Pressure provides an independent physical constraint on phase assignment. To test the pressure fit itself, tungsten diffraction from the gasket was
fitted without using the recorded pressure as a prior and was allowed to
search 0--100~GPa. Using the third-order Birch--Murnaghan equation of state of Dewaele \textit{et al.} \cite{dewaele2004}, the diffraction-derived pressure agreed with the independent ruby measurement to a median absolute difference of 0.72~GPa over the 1036 frames below 20~GPa. At higher pressures the two scales diverged, where both ruby uncertainty under nonhydrostatic conditions and the spatial separation between tungsten in the gasket and ruby in the sample chamber become relevant.

 A separate constrained/unconstrained comparison demonstrates why the experimental pressure prior is retained for routine identification. Without the constraint, 18\% of tungsten fits differed from the recorded pressure by more than 5~GPa, and neon was assigned in more than ten times as many frames, often at pressures far from the measurement conditions. The pressure prior therefore acts as a physical rejection criterion against otherwise plausible positional coincidences.

 \subsection{Computational and manual workload}
The practical outcome is a change in the scale at which a high-pressure series can be analyzed. The original frame-by-frame treatment of this dataset required more than two months to identify useful patterns and separate dominant nonsample contributions. The corresponding SeriesXRD analysis completes overnight. Because the automated result can be inspected through the fitted peaks, assignments, subtracted patterns and residual tracks, the reduction in labor does not require treating the program as a black box. The user can instead begin manual inspection with the small subset of frames and reflections that the series-level analysis identifies as scientifically relevant.

 \section{Scope and limitations}
SeriesXRD is a screening and phase-assignment workflow, not a replacement for crystallographic refinement. Its phase calls depend on the accuracy and validity range of the lattice parameters and equations of state supplied to the library. Published pressure scales can differ, and a phase model should not be extrapolated across a known structural transition. The pressure constraint reduces false matches but can also reject a real assignment if the pressure metadata or equation of state is wrong; both should therefore remain visible to the user.

 The method also assumes that the diffraction signal can be represented adequately by the selected one-dimensional channel and peak profile. Severe texture, very small grain counts, unresolved multiplets or non-pseudo-Voigt peak shapes can bias fitted positions and intensities. In particular, the same azimuthal localization used to identify diamond spots can occur for sample reflections from a coarse-grained or nearly single-crystalline specimen. SeriesXRD tests for this failure mode and can switch fitting channels, but such data remain intrinsically less suited to automated powder-style analysis.

 Unexplained-reflection clustering should be interpreted conservatively. A cluster means that several residual reflections recur and evolve together after the supplied known phases have been removed. It does not establish chemical composition, space group or crystal structure. Those assignments require additional physical information and, where appropriate, indexing and refinement. The advantage of the clustering step is therefore triage: it identifies where coherent unexplained diffraction exists and makes those frames available for targeted analysis.

 Similarly, intensity-share phase fractions are screening quantities because texture, absorption and phase-dependent scattering power are not fully corrected. Quantitative phase fractions and final structural parameters should be obtained through Rietveld refinement, for which SeriesXRD can export the patterns, phase files and instrument information to GSAS-II \cite{toby2013}. These boundaries are deliberate: the software automates the repetitive series-level decisions while leaving crystallographic conclusions to methods designed for them.

 Although developed for diamond-anvil-cell experiments, the underlying workflow is not restricted to pressure as the independent variable. Any diffraction experiment that produces a sequence of related patterns and contains repeated known contributions could in principle use the same approach, provided appropriate metadata and phase models are supplied. The present validation, however, is specifically for high-pressure X-ray diffraction; performance in other experimental geometries should be established separately rather than assumed from the results reported here.

\section{Conclusions}
SeriesXRD addresses a practical bottleneck in modern high-pressure X-ray diffraction: experiments can generate thousands of patterns faster than researchers can determine which frames contain useful sample scattering and which peaks belong to the pressure assembly. The program treats the dataset as a series, combining localized-scattering suppression, background subtraction, peak tracking, pressure-aware phase matching, known-phase subtraction and residual-feature tracking in one reproducible workflow.

The 1288-frame UOTe test demonstrates the intended regime. An approximately 30~nm sample produced detectable diffraction in only a few percent of the measured frames, while diamond, tungsten and pressure-medium signals were present throughout the experiment. SeriesXRD reproduced the sample-bearing frames selected by manual analysis, assigned 90\% of accepted first-pass peaks to known phases, and exposed additional weak reflections after subtraction and re-fitting. The original manual analysis required more than two months; the automated workflow completes overnight.

The principal benefit is therefore not simply faster peak fitting. It is the ability to reduce a large, sparse diffraction series to an auditable set of sample-associated and unexplained reflections that a scientist can inspect directly. For experiments in which manual frame-by-frame analysis is the limiting step, this changes the task from searching the entire dataset to interpreting the small fraction that remains scientifically unresolved.

 \section*{Acknowledgements}
SeriesXRD builds on pyFAI and fabio and interoperates with GSAS-II. We thank their developers and communities.

\section*{Author contributions}
ZR began the project and wrote the early prototype from which SeriesXRD developed, including the initial calibration workflow. ZR also collected the high-pressure series used for the experimental validation at the Advanced Photon Source, together with the accompanying ruby pressure measurements. RM designed and wrote the software described here, implemented the reduction and analysis workflow, ran the validation and benchmarks, and drafted the manuscript. HL helped with the development of the software. SR supervised the work. All authors reviewed and edited the manuscript.

\section*{AI-use disclosure}
AI-based coding assistants were used in developing parts of the software, in writing analysis and evidence-generation scripts, and in drafting portions of this manuscript. All numerical results reported here were regenerated and checked by the authors from the archived scripts, and the authors take responsibility for the content.

 \section*{Funding}
Research at Washington University was supported by the National Science Foundation (NSF) Division of Materials Research Award DMR-2236528. R. M. was supported by the NSF under Grant No. PHY-2547979 (\textit{REU Site: Artificial Intelligence in Modern Physics Research}). %Any opinions, findings, and conclusions or recommendations expressed in this material are those of the authors and do not necessarily reflect the views of the National Science Foundation.

\section*{Disclosure statement}
The authors declare no conflicts of interest.

\section*{Data availability statement}
SeriesXRD is available at \url{https://github.com/RushMaples/SeriesXRD} under the MIT license, from PyPI as \texttt{seriesxrd}, and archived at Zenodo (doi:10.5281/zenodo.21511143). Scripts used to generate the quantitative validation results are distributed with the source. The high-pressure dataset is the subject of a separate study in preparation; a referee-private archive containing the integrated one-dimensional patterns, the complete analysis file and background-corrected text exports is available to the editor and referees on request. Raw detector frames are not included.


\begin{thebibliography}{99}

\bibitem{ashiotis2015}
Ashiotis G, Deschildre A, Nawaz Z, Wright JP, Karkoulis D, Picca FE, Kieffer J. The fast azimuthal integration Python library: pyFAI. J Appl Crystallogr. 2015;48:510--519. doi:10.1107/S1600576715004306.

\bibitem{prescher2015}
Prescher C, Prakapenka VB. DIOPTAS: a program for reduction of two-dimensional X-ray diffraction data and data exploration. High Pressure Res. 2015;35:223--230. doi:10.1080/08957959.2015.1059835.

\bibitem{toby2013}
Toby BH, Von Dreele RB. GSAS-II: the genesis of a modern open-source all-purpose crystallography software package. J Appl Crystallogr. 2013;46:544--549. doi:10.1107/S0021889813003531.

\bibitem{ryan1988}
Ryan CG, Clayton E, Griffin WL, Sie SH, Cousens DR. SNIP, a statistics-sensitive background treatment for the quantitative analysis of PIXE spectra in geoscience applications. Nucl Instrum Methods Phys Res B. 1988;34:396--402. doi:10.1016/0168-583X(88)90063-8.

\bibitem{morhac1997}
Morh\'a\v{c} M, Kliman J, Matou\v{s}ek V, Veselsk\'y M, Turzo I. Background elimination methods for multidimensional coincidence gamma-ray spectra. Nucl Instrum Methods Phys Res A. 1997;401:113--132. doi:10.1016/S0168-9002(97)01023-1.

\bibitem{birch1947}
Birch F. Finite elastic strain of cubic crystals. Phys Rev. 1947;71:809--824. doi:10.1103/PhysRev.71.809.

\bibitem{vinet1987}
Vinet P, Ferrante J, Rose JH, Smith JR. Compressibility of solids. J Geophys Res. 1987;92:9319--9325. doi:10.1029/JB092iB09p09319.

\bibitem{dewaele2004}
Dewaele A, Loubeyre P, Mezouar M. Equations of state of six metals above 94~GPa. Phys Rev B. 2004;70:094112. doi:10.1103/PhysRevB.70.094112.

\end{thebibliography}
\end{document}